\documentclass[aip,reprint]{revtex4-2}
\usepackage[utf8]{inputenc}
\usepackage[T1]{fontenc}
\usepackage{epsfig}
\usepackage{float}
\usepackage{mathptmx}
\usepackage{dcolumn}
\usepackage{bm}
\usepackage{graphicx}
\usepackage{times}
\usepackage{siunitx}
\usepackage{float}
\usepackage{amssymb}
\usepackage{amsmath}
\usepackage{amsfonts}
\usepackage{gensymb}
\usepackage{enumitem} 
\usepackage{setspace}
\usepackage[dvipsnames]{xcolor}
\usepackage{titlesec}
\usepackage{upgreek}
\usepackage{mathtools}
\usepackage[sort&compress]{natbib}

\setcitestyle{open={},close={},numbers,comma,super}

\usepackage[pagebackref=false]{hyperref}
\hypersetup{
    colorlinks  =   true,
    linkcolor   =   BlueViolet,
    citecolor   =   BlueViolet,
    filecolor   =   BlueViolet,
    urlcolor    =   BlueViolet}
\usepackage[capitalise,nameinlink]{cleveref}
\usepackage{hypcap} 
\usepackage{titlesec}
\titleformat{\section}[hang]{\small\bfseries\sffamily}{\thesection.}{0.5em}{\MakeUppercase}
\titlespacing{\section}{0pc}{1.2pc}{0.3pc}
\titlespacing{\subsection}{0pc}{1pc}{0.2pc}

\makeatletter
\renewcommand*{\fnum@figure}{{\normalfont\bfseries \figurename~\thefigure}}
\renewcommand*{\@caption@fignum@sep}{\textbf{ : }}
\makeatother

\begin{document}

\title{An unexplored MBE growth mode reveals new properties of superconducting NbN}

\author{John Wright}
\affiliation{Department of Materials Science and Engineering, Cornell University, Ithaca, New York 14853, USA.}

\author{Celesta Chang}
\affiliation{Department of Physics, Cornell University, Ithaca, New York, 14853, USA}

\author{Dacen Waters}
\affiliation{Department of Physics, Carnegie Mellon University, Pittsburgh, Pennsylvania 15213, USA}

\author{Felix L\"upke}
\affiliation{Department of Physics, Carnegie Mellon University, Pittsburgh, Pennsylvania 15213, USA}

\author{Lucy Raymond}
\affiliation{Department of Materials Science and Engineering, Cornell University, Ithaca, New York, 14853, USA}

\author{Rosalyn Koscica}
\affiliation{Department of Materials Science and Engineering, Cornell University, Ithaca, New York, 14853, USA}

\author{Guru Khalsa}
\affiliation{Department of Materials Science and Engineering, Cornell University, Ithaca, New York, 14853, USA}

\author{Randall Feenstra}
\affiliation{Department of Physics, Carnegie Mellon University, Pittsburgh, Pennsylvania 15213, USA}

\author{David Muller}
\affiliation{Department of Applied and Engineering Physics, Cornell University, Ithaca, New York, 14853, USA}
\affiliation{Kavli Institute for Nanoscale Science, Cornell University, Ithaca, New York, 14853, USA}

\author{Huili G. Xing}
\affiliation{Department of Materials Science and Engineering, Cornell University, Ithaca, New York 14853, USA.}
\affiliation{School of Electrical and Computer Engineering, Cornell University, Ithaca, New York 14853, USA.}
\affiliation{Kavli Institute for Nanoscale Science, Cornell University, Ithaca, New York 14853, USA.}

\author{Debdeep Jena}
\affiliation{Department of Materials Science and Engineering, Cornell University, Ithaca, New York 14853, USA.}
\affiliation{School of Electrical and Computer Engineering, Cornell University, Ithaca, New York 14853, USA.}
\affiliation{Kavli Institute for Nanoscale Science, Cornell University, Ithaca, New York 14853, USA.}

\begin{abstract}
Accessing unexplored crystal growth conditions often reveal new regimes of physical behavior. In this work, performing molecular beam epitaxy growth of the technologically important superconductor NbN at temperatures greater than 1000 ${\degree}$C reveals a growth mode that has not been accessed before. This mode results in persistent RHEED oscillations through the entire growth, resulting in atomically smooth surfaces, normal metal resistivities of $\sim  37$ $\mu\Omega$-cm. We find that the superconducting critical temperature depends strongly on growth temperature, and report a maximum superconducting critical temperature of 15.5 K. Electron-microscopy studies reveal a rich range of crystalline phases that depend on the growth temperature and correlate to the physical properties. Surprisingly, a reversal of the sign of the Hall coefficient from n-type to p-type is observed as the NbN films are cooled, indicating an electronic structure that has not been observed before in this material. In addition to this observation, the crystallinity of the high-temperature epitaxial NbN allows for an ordered Abrikosov vortex lattice to be imaged for the first time in this superconductor.
\end{abstract}
\maketitle


\section{Introduction}
Superconductivity in NbN was discovered in 1941$\cite{Aschermann1941}$. For several years it was was the material with the highest measured superconducting critical temperature at about 17 K. Though materials with higher critical temperatures have since been discovered, NbN is still commonly used in superconducting electronics due in part to the relative ease of thin film deposition, its appreciable critical temperature, its compatibility with standard semiconductor fabrication methods. NbN thin films are used for a wide variety of superconductive electronic devices, including qubits$\cite{Nakamura2011}$, superconducting nanowire single photon detectors (SNSPD)$\cite{Goltsman2001}$ and single flux quantum circuits$\cite{Terai2001}$.
Most commonly, NbN thin films have been created with sputter deposition techniques$\cite{Shoji1992}$, though pulsed laser deposition$\cite{Senapati2006}$, atomic layer desposition $\cite{Ziegler2013}$, chemical vapor deposition$\cite{Hazra2016b}$, and molecular beam epitaxy (MBE)$\cite{Katzer2020, Wang2017}$. Sputter deposition has proven capable of producing NbN films with critical temperatures of 17 K and resistivities as low as 31 $\mu\Omega$-cm. Yet in applications where NbN films thinner than 10 nm are desired, film nucleation, coalescence, and epitaxy become increasingly important.

Recently there is growing interest in integrated heteroepitaxial thin film structures combining superconducting, metallic, insulating, and semiconducting materials to advance quantum computation and communication. For instance, all-epitaxial Josephson junctions have been proposed as a potential method to eliminate the decoherence in qubits caused by amorphous barrier materials$\cite{Simmonds2004}$. We hypothesize that if the precise and ultra-thin layer thicknesses, atomically sharp interfaces, and low defect density characteristic of III-N semiconductor heterostructures grown by MBE could include superconducting transition metal nitrides, novel devices could be realized$\cite{Yan2018, Jena2019}$.

In this work, growth of NbN thin films by MBE at high temperatures is shown to offer advantages in film quality due to the realization of a novel two-dimensional growth mode not previously reported for this material grown by any method. We find a 2D growth mode of NbN for substrate temperatures at or above what is used in most MBE reactors. Films grown in these conditions exhibit positive room temperature Hall coefficients, Hall coefficient reversal as a function of temperature, and the formation of an ordered Abrikosov vortex lattice, properties not seen before in the metallic and superconducting phases of NbN.

\begin{figure*}[!t]
\capstart
\centering\includegraphics[width=1\linewidth]{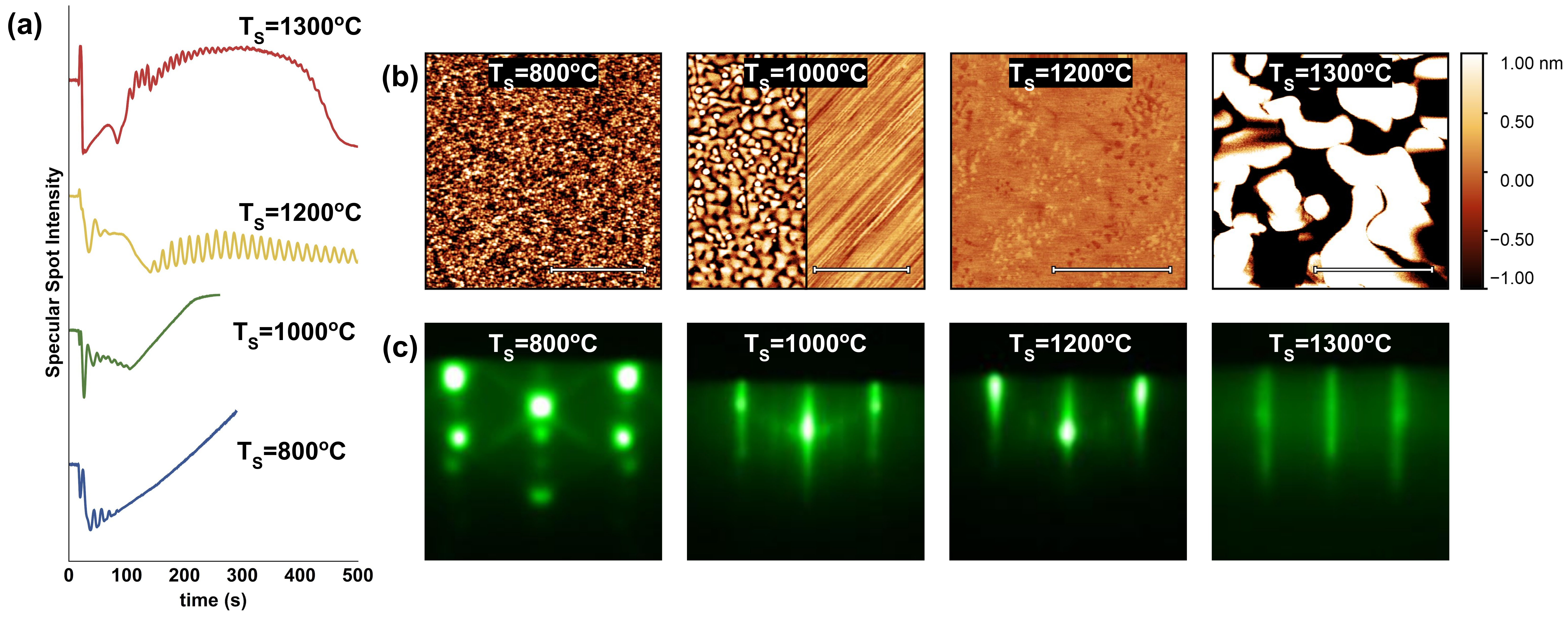}
\caption
{
(a) \textit{In-situ} RHEED specular spot peak intensity as a function of time throughout the first 500s of growth of NbN films on 6H-SiC substrate for different substrate temperatures.
(b) Post-growth AFM surface height maps for NbN films grown at different substrate temperatures. The scale bars correspond to 400 nm.
(c) \textit{In-situ} post-growth RHEED patterns. 
}
\label{fig1}
\end{figure*}

\section{Methods}

\subsection{Thin Film Growth}

NbN thin films were grown on Si-face 6H-SiC substrate by nitrogen plasma-assisted molecular beam epitaxy using a Veeco GENxplor MBE system in which the base pressure is below 10\textsuperscript{-10} Torr. Nb is supplied by an electron-beam evaporator, and Nb flux is measured using an electron impact energy spectroscopy (EIES) system. Film growth is monitored in situ using RHEED. The 6H-SiC substrate was treated by the vendor with a H\textsubscript{2} gas anneal to achieve an atomically terraced surface. Prior to loading into the reactor, substrates were rinsed with solvents followed by rinsing in piranha solution. Immediately prior to loading into the reactor chamber the substrate surface was rinsed with hydrofluoric acid, followed by deionized water to remove the surface native oxide layer. We attempted treating the surface using deposition and thermal desorption of Ga in the MBE reactor chamber to further remove the substrate oxidation layer, a practice which has been utilized for MBE growth of AlN on 6H-SiC\cite{Okumura2012}. However, we found that this practice was correlated with an inability to observe evidence of a 2-dimensional NbN growth mode. A series of NbN samples all approximately 30 nm thick were grown using a range of substrate temperatures. A nitrogen flow rate of 1.95 SCCM and a plasma power of 200 W were used for all growths. The chamber pressure during growth obtained with these conditions was 1.7$\times$10\textsuperscript{-5} Torr; it was observed that the chamber pressure dropped by approximately 10\% as the Nb flux is increased.

The active nitrogen flux for these plasma conditions was determined to be 4.4 atom$\cdot$nm\textsuperscript{-2}$\cdot$s\textsuperscript{-1} by measuring the growth rate of GaN grown in nitrogen limited growth conditions in the same system. For all films presented in this study the flux of active nitrogen during growth exceeds the Nb flux. Growth rates varied slightly between samples, from approximately 0.16 to 0.22 $\si{\angstrom}$/s. These growth rates correspond to a Nb flux between 0.76 and 1.0 atom$\cdot$nm\textsuperscript{-2}$\cdot$s\textsuperscript{-1}.
Substrate temperatures during growth are measured by a thermocouple mounted behind the substrate and are varied between 800 $\degree$C and 1700 $\degree$C. We report the uncalibrated substrate thermocouple temperatures during growth despite our awareness that the substrate thermocouple is not always an accurate measurement of the true surface temperature during growth. However even for a constant substrate thermocouple reading the surface temperature varies during the growth of the NbN film due to increasing absorption of infrared thermal radiation by the high density of conduction electrons in the metallic NbN\cite{ScottKatzer2020}.

\subsection{Crystal Structure and Electronic Properties}
Film surface morphology and roughness were characterized by atomic force microscopy using an Asylum Research Cypher ES system operated in tapping mode.  In-situ phase characterization was performed using a RHEED system operated at 15 kV and 1.5 A. The structural phase and orientation of the films were analyzed using XRD coupled 2$\theta$/$\omega$ and RSM techniques with a Malvern Panalytical Empyrean diffractometer at 45 kV, 40 mA with Cu K$\alpha$1 radiation (1.54057 $\si{\angstrom}$). Film thickness and growth rate were determined using X-ray reflectivity (XRR). Crystal orientation and grain structure were investigated using a Tescan Mira3 SEM-EBSD system. The Nb:N ratio was measured by RBS. 

Further structural analysis was performed by preparing TEM cross-sectional samples using Helios G4-UX Focused Ion Beam (FIB). Prior to ion milling, carbon and platinum were deposited to protect the surface from potential ion-beam damage. The final milling was done at 5kV to reduce redeposition. HAADF-STEM imaging was performed using an aberration-corrected Titan Themis at an acceleration voltage of 300keV. 

Normal metal and superconducting properties were characterized using Van der Pauw resistivity and Hall effect measurements as a function of temperature in a Quantum Design PPMS system. The superconducting properties were further characterized with STM measurements at 4.7 K.

\section{Results}
\subsection{Epitaxial growth and structural properties}

The significant effect of temperature on the growth of the NbN thin films is first discussed. During growth of NbN on 6H-SiC, oscillations of the RHEED specular spot intensity are observed at nucleation, but typically persist only for a time that corresponds to the growth of several NbN layers (Fig. \ref{fig1}(a)). As seen in Fig. \ref{fig1}(c), films grown at 800 $\degree$C exhibit spotty RHEED patterns during and after growth, indicating a 3-dimensional film surface. AFM investigation of the NbN films grown at 800 $\degree$C (Fig. \ref{fig1}(b)) show a smooth surface composed of grains whose diameter on the surface is less than 50 nm. Both the RHEED and AFM data for films grown at 800 $\degree$C are indicative of a 3-dimensional Volmer-Weber (island formation) growth mode.

Films grown at a substrate temperature of 1000 $\degree$C show oscillations of the RHEED specular spot brightness that persisted for slightly more than 100 s (Fig. \ref{fig1}(a)). The RHEED pattern of films grown at 1000 $\degree$C contains a combination of streaky and spotty features (Fig. 1(c)). The surfaces of these films possess some regions with triangular pyramidal grains indicative of 3-dimensional growth of a {1 1 1} oriented cubic film, while other regions possess parallel striped facets across the surface (Fig. \ref{fig1}(b)). The width of each stripe is approximately 10 nm, and the height difference between peak and valley is approximately 0.5 nm. The two morphologies are distributed randomly across the sample surface and can be distinguished through an optical microscope. Striped regions of different orientations meet forming 60$\degree$ angles. This striped morphology appears reliably in several samples grown in these conditions. Using a combination of X-ray diffraction (XRD) and AFM it is determined that the stripes are parallel to the 6H-SiC substrate <1 0 0> direct lattice vector.

Samples grown at a substrate temperature of 1200 $\degree$C show oscillations of the specular RHEED spot brightness (Fig. \ref{fig1}(a)) which persist typically for the growth of at least 10 nm of NbN, though NbN films over 30 nm in thickness have been grown exhibiting RHEED oscillations throughout the entirety of the film growth. We attribute these oscillations to a 2D Frank-van der Merwe (layer-by-layer) growth mode\cite{Braun1999}. The thickness of material grown during a single RHEED oscillation, calculated using the measured film thickness, corresponds well with the 2.51 $\si{\angstrom}$ interplanar lattice spacing of the film measured by XRD.

To our knowledge this is the first time a two-dimensional growth mode has been reported for NbN thin films grown by any method. As shown in Fig. \ref{fig1}(c), for films grown at 1200 $\degree$C the RHEED patterns are streaky with a 2$\times$2 reconstruction pattern. AFM measurements of films grown at 1200 $\degree$C are consistent with layer-by-layer growth (Fig. 1(b)): the surfaces are atomically flat over large regions, with small features above and below the surface whose step heights measure approximately 2.5 $\si{\angstrom}$, corresponding to the interplanar distance of (1 1 1) oriented $\delta$-NbN.

Growths were also attempted at substrate temperatures of 1300$\degree$C, 1500$\degree$C, and 1700$\degree$C. These films all exhibited RHEED patterns that were much dimmer and more diffuse than films grown at lower temperatures (Fig. \ref{fig1}c)), indicating poor crystalline quality. The surface of all samples grown at 1300 $\degree$C or higher had RMS roughness values of several nanometers and showed pits that were roughly as deep as the film thickness (Fig. \ref{fig1}(b)). A variety of nitrogen plasma conditions were attempted at these higher temperatures, but none resulted in films of higher crystal quality or smoother surface morphology.

The metallic and superconducting properties of NbN films depend sensitively on the crystal structure. The Nb-N crystal structure phase diagram is notably complex, containing at least 9 distinct NbN phases\cite{Oya1974, Treece1995a, Babu2019}.  The rock-salt structure $\delta$-NbN phase possesses the highest superconducting critical temperature of all NbN phases and is therefore desired for many superconducting applications\cite{Lengauer1990a}. The primitive cubic Nb\textsubscript{3}N\textsubscript{3} and the tetragonal $\gamma$-Nb\textsubscript{4}N\textsubscript{3}, both structurally similar to $\delta$-NbN, have also been shown to be superconductors with critical temperatures in excess of 10K, though for the structurally similar tetragonal Nb\textsubscript{4}N\textsubscript{5}, no superconductivity has been observed for measured temperatures down to 1.77 K\cite{Oya1974}. The hexagonal close packed $\beta$-Nb\textsubscript{2}N has been reported to have a critical temperature as high as 8.6 K\cite{Gavaler1969}, and the two related hexagonal phases $\delta$'-NbN and $\epsilon$-NbN, have been reported to not transition to the superconducting state above 1.77 K\cite{Babu2019, Oya1974}, though recent reports claim higher transition temperatures for the $\epsilon$-NbN phase\cite{Zou2016}.

\begin{figure}[htp]
\capstart
\centering\includegraphics[width=1\linewidth]{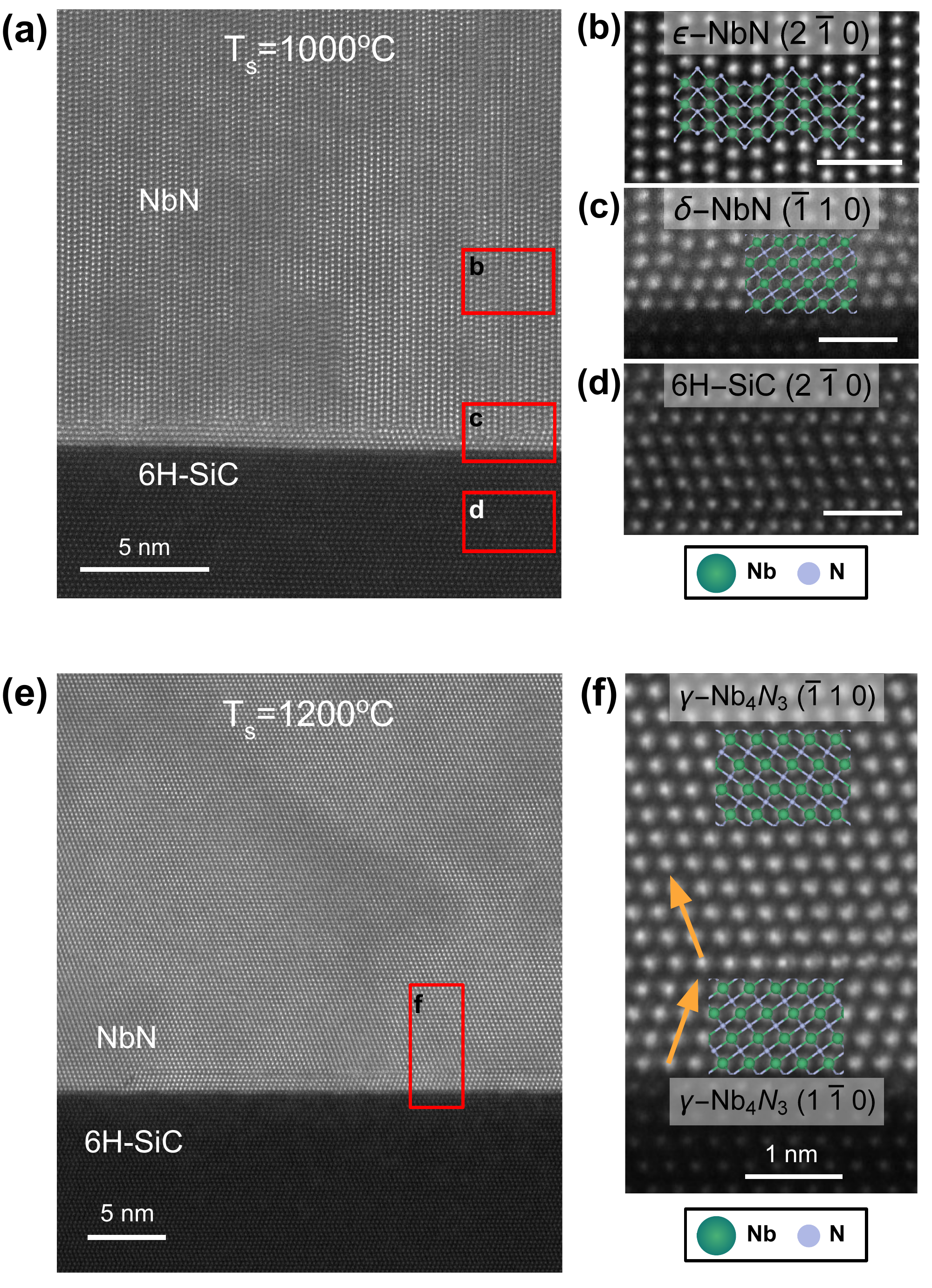}
\caption
{
(a)-(d) HAADF-STEM image of an NbN film grown at 1000 $\degree$C shows a transition from cubic $\delta$-NbN to hexagonal $\epsilon$-NbN a few unit cells away from the substrate. Scale bars in (b)-(d) correspond to 1nm. (e) \& (f) HAADF-STEM image of a NbN film grown at 1200 $\degree$C reveals a structure consistent with both the $\delta$-NbN phase and the closely related $\gamma$-Nb\textsubscript{4}N\textsubscript{3}. Near the 6H-SiC/NbN interface are several layers of NbN that have a lattice orientation rotated 60$\degree$ about the growth axis relative to the orientation that is dominant in this image.
}
\label{fig2}
\end{figure}

\begin{table*}[!t]
 \capstart
 \begin{tabular}{c c c c c c c c} 
 \hline\hline
 T\textsubscript{s} ($\degree$C) ~&~ Phase(s) ~&~ T\textsubscript{c} (K) ~&~ AFM R\textsubscript{RMS} (nm) ~&~ RRR ~&~ $\rho$@300 K ($\mu\Omega\cdot$cm) ~&~ R\textsubscript{H}@300 K ($\mu\Omega\cdot$cm$\cdot$T\textsuperscript{-1}) ~&~ Nb:N ratio\\ [0.5ex] 
 \hline
 800 & $\delta$-NbN & 10.55 & 0.61 & 0.45 & 207.03 & -0.0066 & 1.08$\pm$0.08\\ 
 1000 & $\delta$-NbN/$\epsilon$-NbN & 15.5 & 0.79/0.21 & 1.39 & 79.40 & 0.079 & 1.12$\pm$0.09\\
 1200 & $\gamma$-Nb\textsubscript{4}N\textsubscript{3} & 6.98 & 0.11 & 4.68 & 37.18 & 0.0039 & 1.15$\pm$0.09\\
 1300 & Unidentified & <4 & 6.53 & 1.3 & 56.17 & - & -\\
 \hline\hline
\end{tabular}
\caption
{
Properties of 30 nm NbN films grown on 6H-SiC.
}
\label{table1}
\end{table*}

We have used the combination of XRD, RHEED, and scanning transmission electron microscopy (STEM) to study the phase of the NbN films. The crystal phase(s) of each film is summarized in Table \ref{table1}. To determine the crystal structure of the NbN material grown at 1000 $\degree$C, high angle annular dark field (HAADF-STEM) measurement was performed on a region with the striped surface morphology shown in Fig. 1(b). This measurement (Fig. 2(a)) indicates that the striped regions are composed predominantly of hexagonal $\epsilon$-NbN, with the growth-axis orientation $\epsilon$-NbN [1 2 0] // 6H-SiC [0 0 1] and the in-plane orientation $\epsilon$-NbN [0 0 1] // 6H-SiC [1 0 0].  Therefore, the c-axis of the hexagonal NbN film is orthogonal to the c-axis of the substrate. Several layers of cubic $\delta$-NbN are observed near the NbN/6H-SiC interface, after which the film transitions to the hexagonal $\epsilon$-NbN phase. This thin nucleation layer of $\delta$-NbN was observed in all locations imaged. The regions in which the striped morphology did not appear were determined using XRD to correspond to the cubic $\delta$-NbN phase. It is unclear why only some regions of the film underwent a structural transition from $\delta$- to $\epsilon$-NbN, and why it only occurs at this substrate temperature.

HAADF-STEM images of the NbN film grown at 1200 $\degree$C in Fig. \ref{fig2}(e)-(f) reveal a crystal structure consistent with cubic $\delta$-NbN. Several related phases, such as $\gamma$-Nb\textsubscript{4}N\textsubscript{3} in which the a and c lattice parameters differ by around 1\%, are difficult to distinguish from $\delta$-NbN based only on cross-sectional STEM images. However, XRD measurements provide evidence that the structure of the film is the tetragonal $\gamma$-Nb\textsubscript{4}N\textsubscript{3} phase, as discussed later. The first few layers above the substrate interface are observed to possess a lattice orientation rotated 60$\degree$ about the growth axis from the orientation that is dominant in this image. Given the 6-fold symmetry of 6H-SiC, both orientations of the NbN lattice are equivalent with respect to the substrate lattice.

\begin{figure}[htp]
\capstart
\centering\includegraphics[width=1\linewidth]{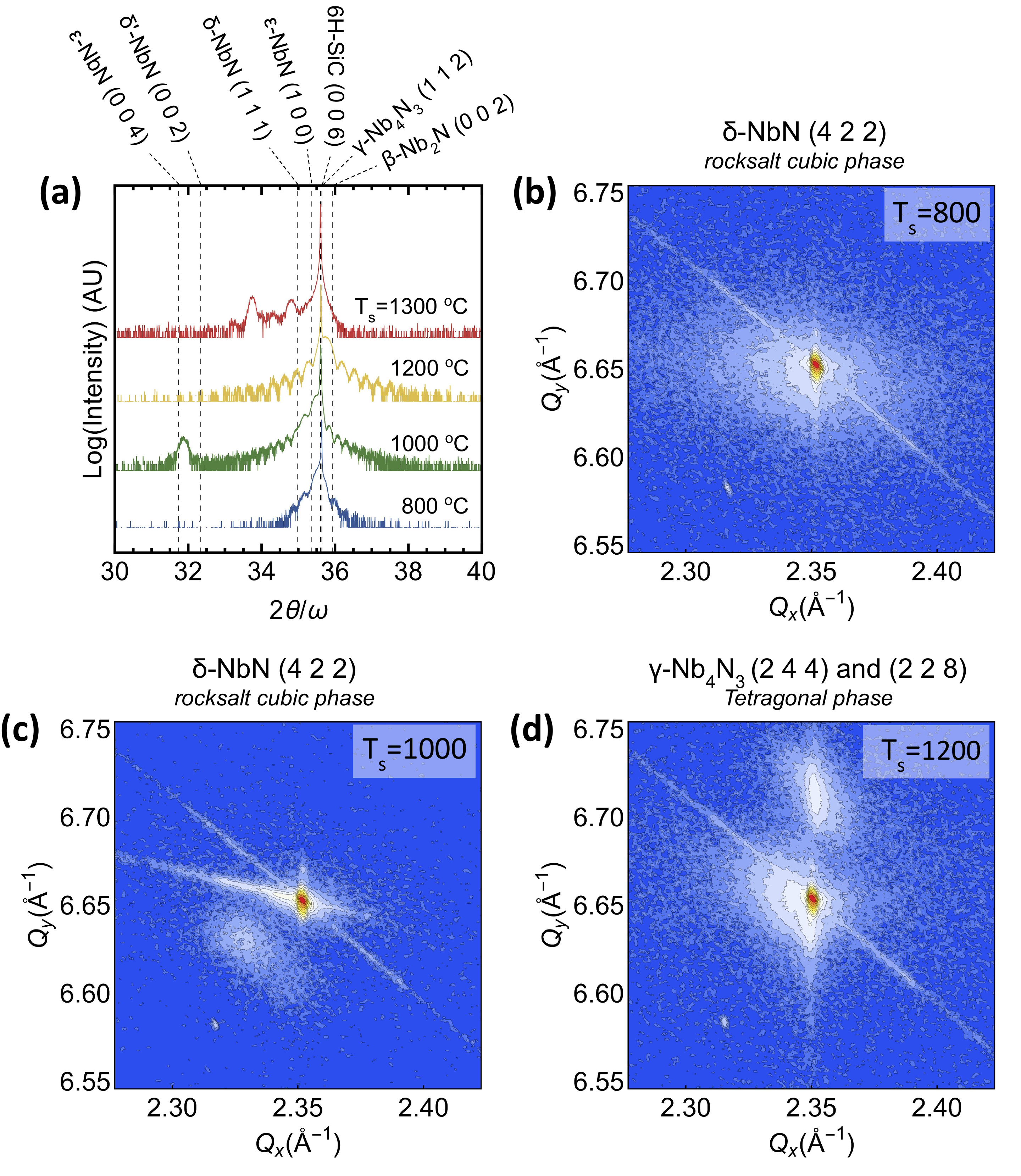}
\caption
{
(a) Symmetric 2$\theta$/$\omega$ XRD scans of 30 nm NbN thin films grown on 6H-SiC substrate. (b)-(d) XRD reciprocal space maps of NbN films grown at (b) 800 $\degree$C, (c) 1000 $\degree$C, and (d) 1200 $\degree$C. The region mapped is centered on the 6H-SiC (1 0 16) RLP. The expected epitaxial relationship of $\delta$-NbN on 6H-SiC places the $\delta$-NbN (4 2 2) RLP near the 6H-SiC (1 0 16) RLP.
}
\label{fig3}
\end{figure}

Based on XRD 2$\theta$/$\omega$ symmetric coupled scans (Fig. \ref{fig3}(a)) we confirm the presence of NbN phases with out-of-plane lattice parameters similar to that of the 6H-SiC substrate for each film. However, given the large number of NbN phases with similar peak locations and the compositional variation possible within each phase\cite{Lengauer1990a}, these measurements of the out-of-plane film lattice parameter alone are insufficient to identify phases uniquely. To further characterize the crystal phase in each film we utilize asymmetric reciprocal space maps in the vicinity of reciprocal lattice points (RLP) of expected NbN phases. Fig. \ref{fig3} shows that all three films exhibit a film peak near the SiC (1 0 16) peak. By mapping a variety of regions of reciprocal space and detecting only peaks that correspond to $\delta$-NbN we conclude that the film grown at 800 $\degree$C is entirely $\delta$-NbN. The $\delta$-NbN phase grows with the growth-axis orientation $\delta$-NbN [1 1 1]//SiC [0 0 1] and the in-plane orientations $\delta$-NbN [1 -1 0]//SiC [1 0 0] and $\delta$-NbN [-1 1 0]//SiC [1 0 0]. Due to the 6-fold symmetry of the substrate and the 3-fold symmetry of the $\delta$-NbN about the growth axis, two in-plane orientations of the $\delta$-NbN lattice are possible, differing by a 60$\degree$ rotation about the growth axis. RHEED and XRD reciprocal space mapping (RSM) confirm the presence of both $\delta$-NbN orientations. Therefore, it is concluded that the $\delta$-NbN films grown on 6H-SiC are not single crystal but composed of cubic twin grains.

As seen in Fig. \ref{fig3}(d), the NbN film grown at 1200 $\degree$C exhibits a splitting of the NbN RLP such that there are 2 film peaks near the 6H-SiC (1 0 16) RLP. This splitting is evidence of tetragonal distortions of the rock-salt $\delta$-NbN structure. This evidence in combination with the RHEED pattern allows us to conclude that the film possesses the $\gamma$-Nb\textsubscript{4}N\textsubscript{3} structure. A single-orientation $\gamma$-Nb\textsubscript{4}N\textsubscript{3} film would not exhibit two reciprocal lattice points at this location. However, for a film composed of tetragonal grains that are related by 120 $\degree$C rotation about the growth-axis, both the (2 4 4) and (2 2 8) reciprocal lattice points will be visible at this location. The grain structure can therefore be understood as the combined result of cubic twinning and tetragonal distortions of different lattice vectors in different regions of the film, producing 6 different orientations of the tetragonal structure, all of which are observed in the XRD analysis.

For both the film grown at 800 $\degree$C and 1200 $\degree$C, the RSM (Fig. \ref{fig3}b-d) indicates that the in-plane lattice constant of the NbN is identical to that of the 6H-SiC, indicating that the films are strained. This strain requires a rhombohedral distortion to the cubic or tetragonal structures.

In addition to phase characterization, the Nb:N ratio of samples grown at 800 $\degree$C, 1000 $\degree$C, and 1200 $\degree$C was measured using Rutherford backscattering spectrometry (RBS) (Table I). All the films measured had Nb:N ratios greater than 1. The highest Nb:N ratio measured was 1.15$\pm$0.09 for the sample grown at 1200 $\degree$C, with the variation in Nb:N ratio between phases being within the uncertainty of the measurement.

\begin{figure}[htp]
\capstart
\centering\includegraphics[width=1\linewidth]{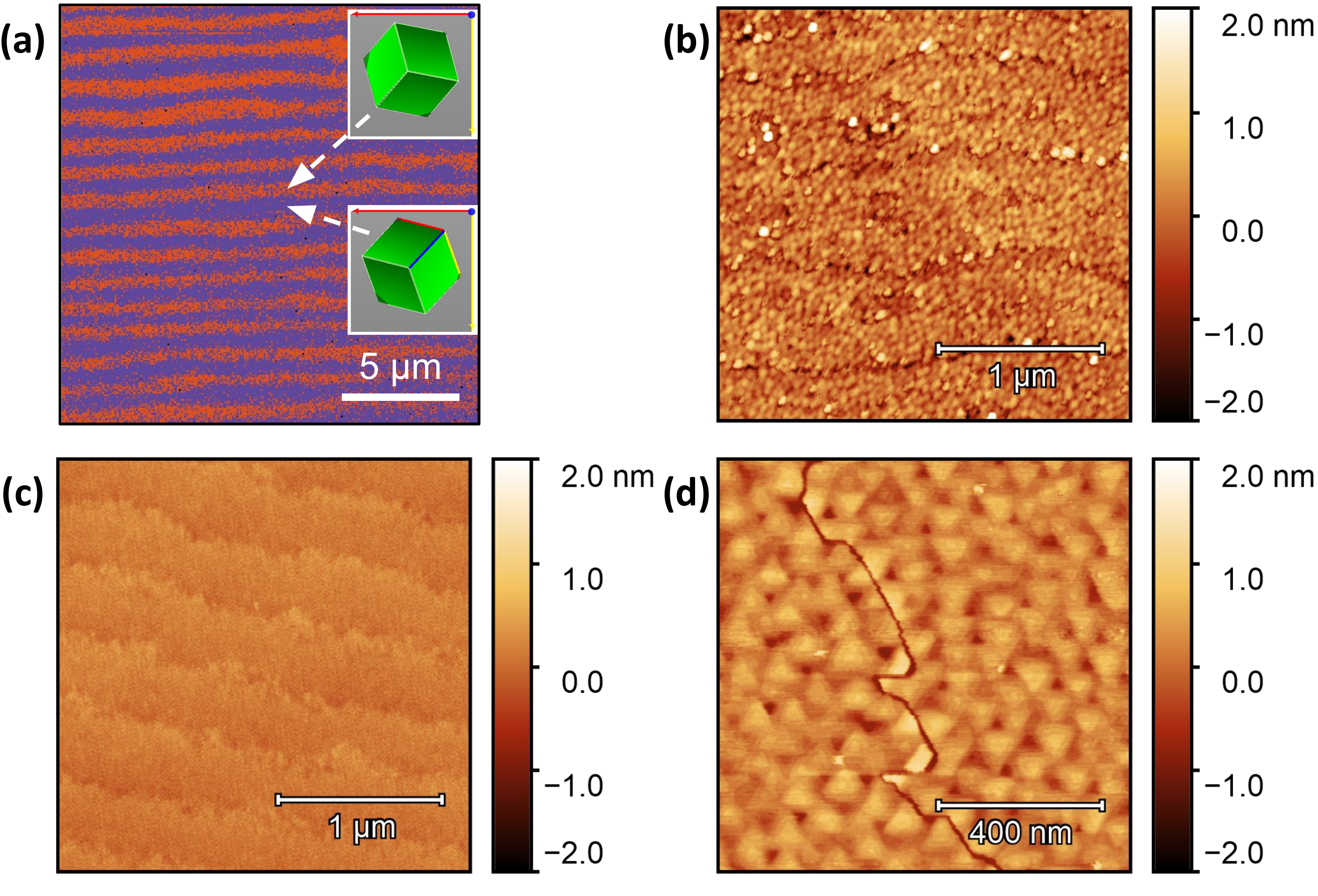}
\caption
{
(a) The EBSD map of a 100 nm NbN film reveals that the cubic twin grain boundaries form parallel stripes across the surface with width of approximately 550 nm. (b) AFM surface mapping of the same NbN film shows a series of parallel depressions whose separation and orientation correspond with the grain boundaries observed in the EBSD map. (c) A region of the 6H-SiC substrate that was shadowed during growth shows atomic terraces whose separation and orientation do not accurately match the grain boundaries orientation and separation. (d) AFM mapping of a NbN sample grown at 1200 $\degree$C at the site of a cubic twin grain boundary shows there is a depression at the site of the boundary approximately 1 nm deep.
}
\label{fig4}
\end{figure}

To characterize the size and distribution of grains within the films, we performed lattice orientation mapping using scanning electron microscopy (SEM) based electron backscatter diffraction (EBSD). A 100 nm pure $\delta$-NbN film grown at 800 $\degree$C was used for EBSD analysis. The EBSD map (Fig. \ref{fig4}(a)) shows the interesting feature that the grains of different $\delta$-NbN orientations form an alternating striped structure. In this film the grain width averages 548$\pm$52 nm. AFM mapping (Fig \ref{fig4}(b)) of the same sample reveals parallel depressions with the same orientation and separation as the boundaries between grains observed in the EBSD map, and we therefore ascribe the depressions to the grain boundaries. A similar grain boundary is shown in Fig. \ref{fig4}(d) for a film grown at 1200 $\degree$C. In this image we observe triangular pyramidal surface features that rotate by 60$\degree$ on either side of the boundary, providing further evidence that the NbN lattice possesses different orientations on either side of the surface depression.

We hypothesized that the atomic steps of the 6H-SiC substrate act as nucleation sites for the NbN film and therefore play a role in the formation of the observed grain structure. To test this hypothesis, we performed AFM mapping of a region of the substrate which was shadowed during growth of the film shown in Fig. 4(a)-(b). We observe atomic terraces with step-heights of 2.50$\pm$0.04 $\si{\angstrom}$, which corresponds well with the height of one sixth of the 6H-SiC c-axis unit cell (i.e. the height of a basal plane atomic layer). We find that the terrace widths average 407$\pm$38 nm and are not oriented along the same direction as the grain boundaries observed by EBSD and AFM. Therefore, this observation does not establish correlation between the substrate atomic terraces and the lamellar grain boundaries.

\subsection{Metallic conduction and Hall effect reversal}

The electrical properties of the films were characterized using temperature dependent Van der Pauw resistivity measurements. Low values of film resistivity are desirable for metallic applications of epitaxial NbN thin films\cite{Miller2020}. We find that the resistivity of NbN samples grown at different temperatures varies over a large range as seen in Table I, with the minimum room-temperature resistivity of 37 $\mu\Omega\cdot$cm  observed for a film grown at 1200 $\degree$C. The highest room temperature resistivity of 207 $\mu\Omega\cdot$cm was observed in a film grown at 800 $\degree$C. The lowest resistivity value previously reported for NbN thin films is 31.2 $\mu\Omega\cdot$cm for a film 100 nm thick grown by reactive RF-magnetron sputtering\cite{Shoji1992}, indicating that further reduction of film resistivity for MBE grown NbN may still be possible.

\begin{figure}[h]
\capstart
\centering\includegraphics[width=1\linewidth]{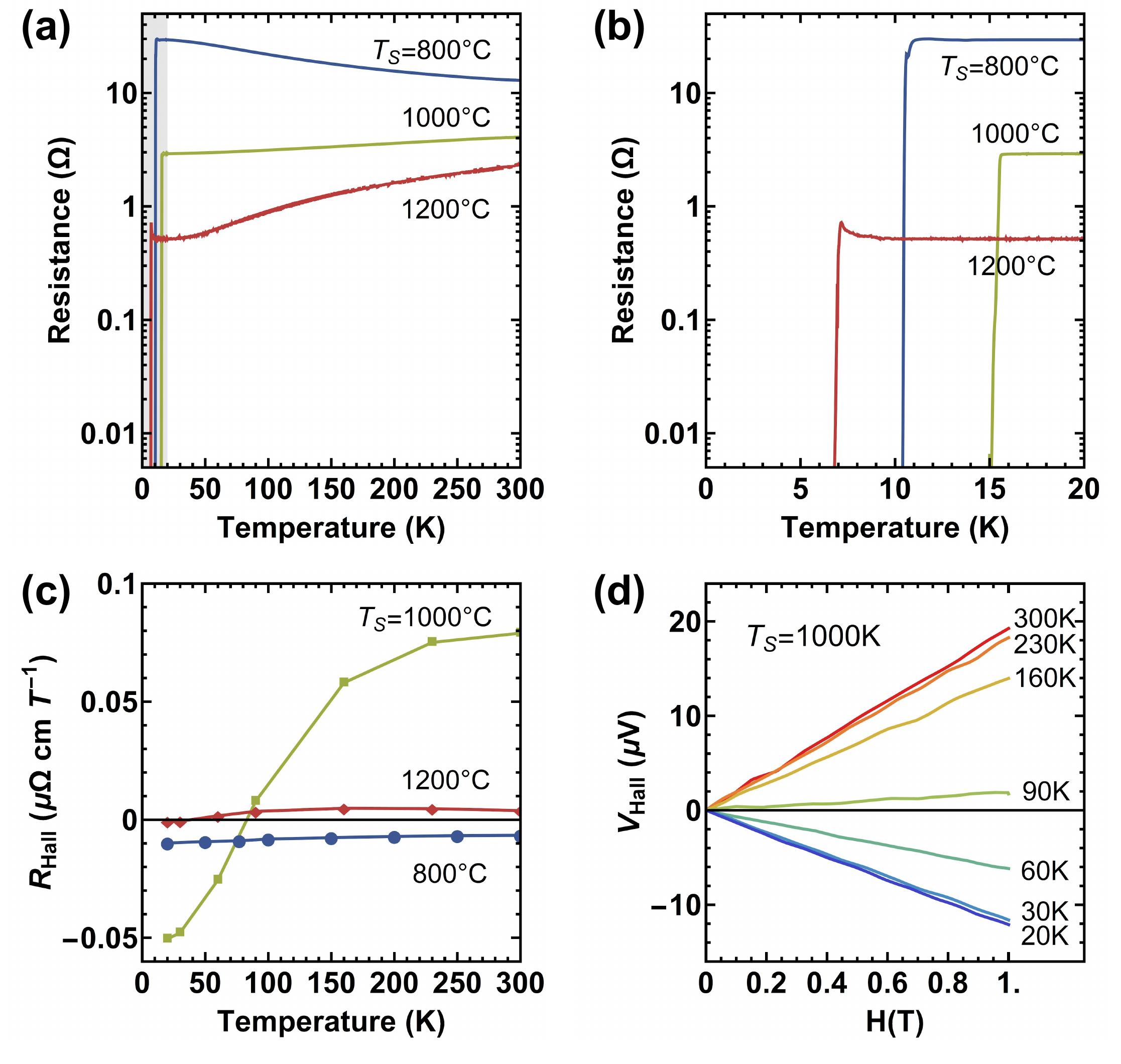}
\caption
{
(a) Normal state resistance versus temperature of NbN thin films. (b) Resistance versus temperature in the range of the transition to the zero-resistance state. (c) Hall resistance versus temperature. (d) Hall voltage versus field for a single NbN film grown at 1000 $\degree$C.
}
\label{fig5}
\end{figure}

The residual resistivity ratio (RRR), defined here as the ratio between the resistance at 300 K and the resistance at 20 K is used to compare the temperature dependence of resistivity of different films. As seen in Fig. \ref{fig5}(a) and Table I, the RRR varies directly with the growth temperature, starting at a value of 0.45 for the film grown at 800 $\degree$C and increasing to 4.7 for the film grown at 1200 $\degree$C. The variation of RRR from values greater than one to values less than one has been reported previously in sputtered NbN films by Keskar et al., who found the RRR decreased as the partial pressure of nitrogen during sputtering increased at a constant substrate growth temperature\cite{Keskar1971}.

To further investigate the nature of conductivity of each film, Hall effect measurements of the NbN films were performed at temperatures ranging from 300 K to 20 K by sweeping the magnetic field from -1 T to +1 T. The Hall voltage is confirmed to vary linearly with magnetic field over the range of magnetic fields tested. Fig. \ref{fig5}(c) shows the Hall coefficients as a function of temperature for the three NbN films grown at different substrate temperatures. For the films grown at 1000 $\degree$C and 1200 $\degree$C the Hall coefficient is positive at room temperature and becomes negative at temperatures less than approximately 85 K for the film grown at 1000 $\degree$C (Fig. \ref{fig5}(d)) and less than approximately 50 K for the film grown at 1200 $\degree$C. For the sample grown at 800 $\degree$C, the Hall coefficient is negative at all temperatures measured. Such a sign change in the Hall coefficient as a function of temperature has been observed in other single crystal metals with complex Fermi surfaces, such as indium\cite{Ozimek1978}. Previous measurements of temperature dependent Hall coefficient of epitaxial NbN have observed negative Hall coefficients at all measurement temperatures\cite{Chand2009}. To our knowledge this is the first report of the reversal of the Hall coefficient in NbN and the first report of a positive Hall coefficient at room temperature in NbN. This observation indicates a richer range of n- and p-type regions of the Fermi surface of NbN. It is necessary in the future to perform band structure measurements and 1st-principle calculations to explain this surprising observation, which is suggested as a future work.

\subsection{Superconductivity and Abrikosov vortex lattice}

The transition to the superconducting state is determined using Van der Pauw resistance measurements as a function of temperature (Fig \ref{fig5}(b)). The highest superconducting transition temperature observed was 15.5 K for a sample grown at 1000 $\degree$C. For comparison, the highest reported transition temperature for $\delta$-NbN thin films is approximately 17 K\cite{Lengauer1990a}. We note that the sample from this work with the highest transition temperature is a mixture of $\delta$-NbN and $\epsilon$-NbN, and therefore the superconducting critical temperature of different regions of the film likely differs. The NbN samples grown at 800 $\degree$C, which were determined to be purely $\delta$-NbN, exhibited a lower critical temperature of 10.6 K. We therefore conclude that the growth of a phase pure $\delta$-NbN film is neither necessary nor sufficient to achieve a film with high superconducting critical temperature. Samples grown at 1300 $\degree$C show no transition to the superconducting state above 4 K.

\begin{figure}[htp]
\capstart
\centering\includegraphics[width=1\linewidth]{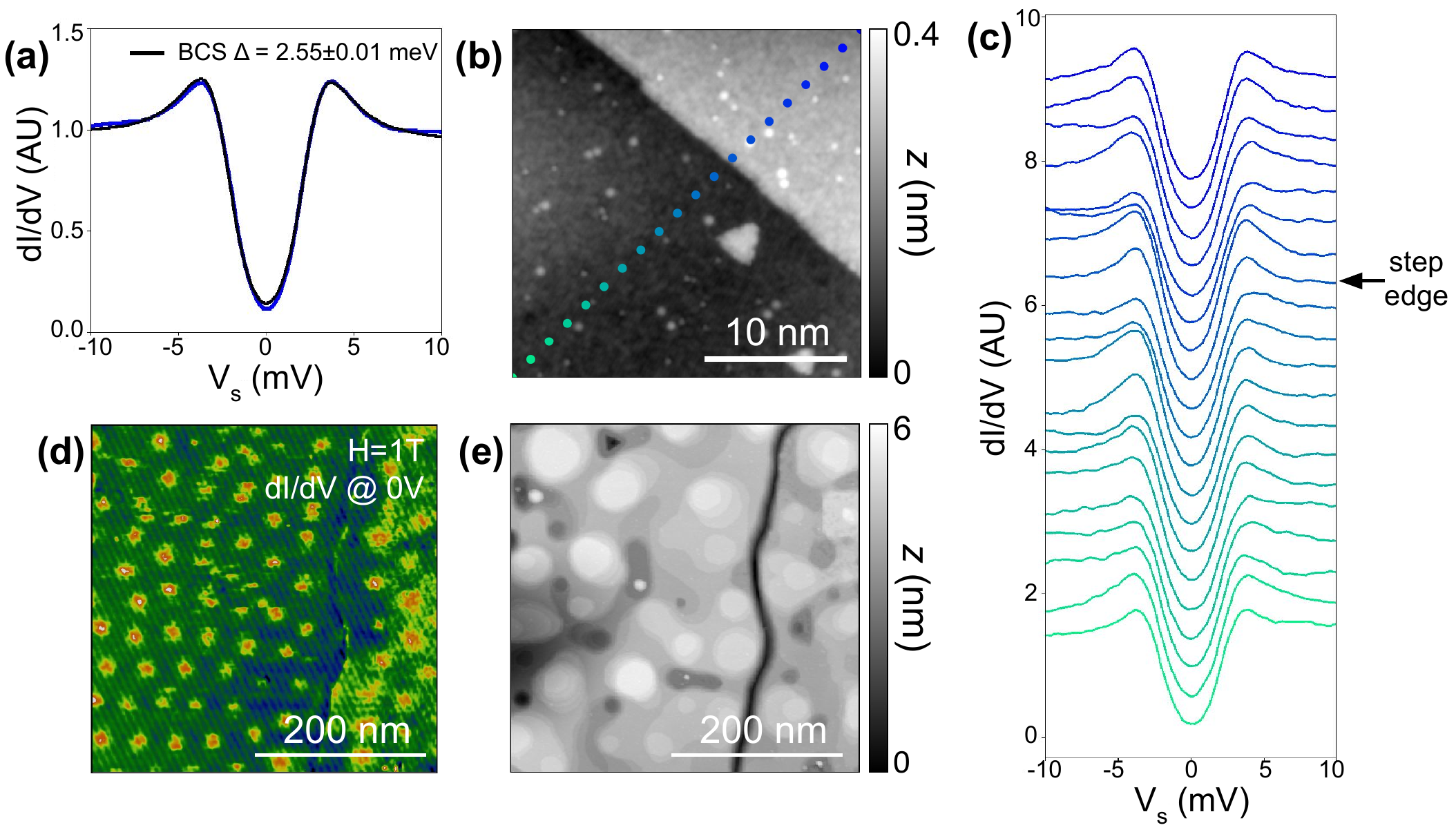}
\caption
{
(a)-(d) Scanning tunneling microscope differential conductance measurements at 4.7 K of a NbN film grown at T\textsubscript{s}=1200 $\degree$C: (a) differential conductance versus voltage is well fit using the BCS density of states yielding the energy gap of the NbN film. (b) Surface topography map and (c) differential conductance waterfall plot shows a highly uniform superconducting energy gap across the sample surface. (d) Differential conductance map and (e) surface topography map in the presence of a 1 T magnetic field shows an Abrikosov vortex lattice that is distorted in the vicinity of a cubic twinning grain boundary. Sample bias was 0.1 V and constant current was 10 pA for both (b) and (e), and modulation voltage was 0.5 mV peak-to-peak for (a), (c), and (d).
}
\label{fig6}
\end{figure}

To further investigate properties of the superconducting state of the NbN films, scanning tunneling microscopy (STM) measurements were performed at 4.7 K for a single NbN film grown at 1200 $\degree$C. Differential conductance measurements, shown in Fig. \ref{fig6}(a), are used to characterize the superconducting energy gap of the film. The superconducting energy gap is determined by modelling the differential conductance based on the BCS density of states\cite{Lupke2020}, yielding an energy gap of $\Delta$=2.55$\pm$0.01 meV. Fig. \ref{fig6}(b) shows the surface topography near an atomic step and Fig. \ref{fig6}(c) shows a waterfall plot of the differential conductance across the region imaged in Fig 6(b). We observe that the superconducting energy gap is uniform across the sample surface, including across an atomic step and a surface defect that appears in the topography scan.

Fig. \ref{fig6}(d) shows a differential conductance map at 0 V in the presence of a magnetic field of 1 T, enabling us to image the presence of Abrikosov vortices. As is apparent in Fig. 6(d) the Abrikosov vortices form a triangular lattice, an indication of the lack of strong pinning of vortices at the sites of material defects. Previous reports of STM analysis of epitaxial NbN thin films did not observe the formation of an ordered vortex lattice\cite{Wang2017}. Fig \ref{fig6}(e) shows the surface topography of the region imaged in Fig \ref{fig6}(d), revealing a depression in the surface that we attribute to a grain boundary between cubic twin grains due to the similarity of the feature to that shown in Fig. \ref{fig4}(d). It is apparent in Fig. \ref{fig6}(d) that the Abrikosov vortex lattice near the cubic twin domain boundary is distorted, indicating that the domain boundary affects the local superconducting state.

\section{Discussion}
We have shown in this study that the growth mode, morphology, phase, and electronic properties of NbN thin films grown by MBE are strongly influenced by substrate growth temperature. With this in mind we note that achieving uniform and controlled film properties is complicated by the fact for a constant substrate thermocouple temperature the film surface temperature increases during growth as the NbN film itself increases infrared absorption from the substrate heater\cite{ScottKatzer2020}.

Previous work on the Nb-N phase diagram has predicted that the $\delta$-NbN phase is thermodynamically stable only at temperatures exceeding 1300 $\degree$C\cite{Levin1994}. It is therefore interesting to note that in the highly non-equilibrium conditions of MBE, films of pure $\delta$-NbN phase appear at substrate temperatures much below this, and that no $\delta$-NbN films have been grown above 1300 $\degree$C despite repeated attempts with a variety of nitrogen and Nb flux ratios and growth rates.

The reversal of the Hall coefficient as a function of temperature in a NbN thin film is a new and surprising observation. Films grown at 800 $\degree$C showed negative Hall coefficient at all temperatures, and those grown at 1000 $\degree$C and 1200 $\degree$C showed positive Hall coefficient at room temperature and negative Hall coefficient at temperatures below 85 K and 50 K respectively. We have demonstrated that films grown at different temperatures differ in growth mode, grain size, crystal phase, and potentially Nb:N ratio, and therefore are currently unable to simply attribute differences in electronic properties between the films to any of the other observed differences between the films. A combination of band structure measurement and theory of the magnetoconductivity and Hall effect in single crystal NbN is needed to explain the surprising Hall effect reversal. The observation of an ordered Abrikosov vortex lattices indicates the feasibility of high-quality epi-NbN for flux-based classical and quantum computation applications.

Although layer-by-layer growth of NbN has been achieved, removing all grain boundaries from the films has not, due to the symmetry mismatch between the substrate and the film. The result is a film composed of single orientation grains approximately 500 nm wide and at least 100 $\mu$m long. Initial evidence from STM measurements indicates that these grain boundaries do affect the superconducting state. The importance of these boundaries will ultimately depend on the intended application of the film. We note that for some applications the grains are large enough and the boundaries are easy enough to locate that it should be possible to fabricate electronic devices on individual single orientation superconducting regions.

\section*{Acknowledgments}
This work used the CNF, CCMR and CESI Shared Facilities partly supported by the NSF NNCI program (ECCS-1542081), MRSEC program (DMR-1719875) and MRI DMR-1338010 and Kavli Institute at Cornell (KIC). We acknowledge funding support from the Office of Naval Research, monitored by P. Maki under award numbers N00014-20-1-2176 and N00014-17-1-2414.
 
\section*{Author contributions}J. Wright prepared the films and performed the XRD, AFM, EBSD, and temperature dependent transport measurements. C. Chang and D. Muller performed the STEM measurements. D. Waters, F. Lüpke, and R. Feenstra performed the STM measurements. L. Raymond assisted with XRD measurements. R. Koscica assisted with Hall effect measurements. G. Khalsa, D. Jena, and H.G. Xing coordinated the work and provided experimental and theoretical guidance. J. Wright and D. Jena wrote the paper.
 
\section*{Competing interests}
The authors declare that they have no conflict of interest.

\section*{Data availability }
The data that support the findings of this study are available from the corresponding author upon reasonable request.
\def\bibsection{\section*{References}}
\bibliography{My_reference}
\end{document}